\documentclass[aps,prb,twocolumn,groupedaddress,showpacs,showkeys,floatfix]{revtex4}
\usepackage{amsmath}
\usepackage{amsfonts}
\usepackage{amssymb}
\usepackage{graphicx}

\begin{document}


\title{Nonlinear conductance in molecular devices:
   molecular length dependence}

\author{\v Z.~Crljen}
\email[]{crljen@irb.hr}
\affiliation{R. Bo\v skovi\' c Institute, P.O. Box 180, 10002 Zagreb,
     Croatia}
\author{A.~Grigoriev}
\author{G.~Wendin}
\affiliation{Department of Microtecnology and Nanoscience, MC2,
    Chalmers University of Technology, SE-41296, G\" oteborg, Sweden}
\author{K.~Stokbro}
\affiliation{Nanoscience Center, Copenhagen University, Universitetsparken 5d
DK-2100 Copenhagen east, Denmark}

\date{August 12, 2004}

\begin{abstract}

We theoretically study the electronic transport in the monolayer of
dithiolated phenylene vinylene oligomeres coupled to the (111) surfaces of gold
electrodes. We use non-equilibrium Green functions (NEGF) and density functional
theory(DFT) implemented in the
TranSIESTA package to obtain a full ab initio self-consistent description of 
the transport current through the molecular 
nanostructure with different electrochemical bias potentials. The calculated 
current-voltage characteristics (IVC) of the systems for the same contact geometry 
have shown a systematic decrease of 
the conductivity with the increased length of the molecules. We analyze the results
in terms of transmission eigenchannels and find that besides the delocalization of
molecular orbitals the distance between gold electrodes also determines the transport 
properties.

\end{abstract}


\maketitle

\section{\label{introduction}Introduction}

Since the pioneering work on electron transport through molecules by Aviram and
Ratner\cite{AvRat} there has been  remarkable progress in the field of
molecular electronics. With the advances of experimental tools and techniques,
various molecular devices have shown a number of interesting
effects: negative differential resistance,\cite{Chen} switching properties,\cite{Collier}
memory effects.\cite{ChRe}
A number of measurements have been made where the electrical current has been driven through
monolayer molecular films between metallic electrodes, including break junction, \cite{Kerg}
and evaporated electrode\cite {Reed} experiments. The connection of a single molecule
to conducting
electrodes has been achieved using STM\cite{Aviram} and break 
junction\cite{Reed,Reichert}
techniques. Owing to their stability the most common systems considered were 
thiol-ended organic molecules on gold electrodes.

The current-voltage measurement by Reed et al.\cite{Reed} of dithiol benzene molecule
(dtb) onto gold electrodes in break-junction experiments showed an
apparent gap of about 0.7 V. It raised the question of its nature: Whether it was a
Coulomb gap or it reflected the mismatch between the contact Fermi level and the
lowest unoccupied molecular orbital. It was viewed as a question of bonding
between the molecule and the electrodes.

In general, in the nonbonding, and even weak-bonding situations one expects no coherent
transport, but instead a sequential charging
of the molecule and Coulomb blockade effects.
Recently Kubatkin et al.\cite{Kubatkin} have demonstrated a Coulomb blockade and different 
charge states in the junction with weakly coupled (physisorbed) OPV5 molecule onto gold
contacts. 

In the strong-bonding case, a coherent transport is dominant and the staircaselike
structure could appear owing to the succession of scattering
resonances entering the bias window.\cite{Kerg}
In many junctions no apparent gap was noticed. Instead, linear current-voltage 
curves were reported at low bias for strongly bonded thiolate
molecules.
There are a number of possible transport mechanisms that give rise to linear IVC. In the
evaporation electrode technique, for example, grainy
surfaces can form with the likely protrusion of
electrode atoms in between the molecules of the monolayer, thus spoiling the surface order and
the dominant conduction mechanism appears to be hopping with the 
characteristic energy scale of 10-100 meV.\cite{Zhitenev}

In order to explain experimental IVCs, the dithiol-benzene molecule coupled to
Au(111) surfaces was a kind of model system for theoretical studies. After the first 
semiempirical investigations,\cite{Yaliraki,Derosa} ab initio calculations within the
framework of density functional theory were performed. The conductance was found to be more than
an order of magnitude larger than in experiment,\cite{Ratner,DiVentra} despite the 
assumption of unequal bonding
of the molecule to the electrode surfaces.\cite{dtb_Au}

In the theoretical model of molecular junction with parallel perfect electrode surfaces, it 
came natural that the
electronic structure of the system provided a finite density of states at the Fermi level of the
coupled system, ruling out the appearance of the gap in the I-V curves.
The theoretically obtained conductances were as a rule more than an
order of magnitude larger compared with the measured ones. The difference was related to the
varying strengths of the sulfur-gold bonds and to the adsorption
sites. Interestingly, for longer molecules, quantitative agreement of theoretical
considerations\cite{Mujica,Heurich,sidegr}
is considerably improved when compared with the measurements.\cite{Cui,Reichert}

The transmission properties of the systems are determined by the electronic structure of
the combined molecule and electrode systems. The distance between the electrodes
has an important role in the conduction of the junction.
In simple terms, the metal-molecule-metal system can be viewed as a finite quantum well
that sets up in between the metal surfaces. The direct tunneling between the electrodes,
despite the possible electrode-induced gap states in the metal-air-metal junctions,
is considered to be too small to contribute considerably to the current of the typical size
  molecular junctions.\cite{Kerg}

With the molecule bridging the interelectrode gap, 
a possibility for the molecule-induced interelectrode gap states, or resonances appears.
Those states will have an important impact on the transport properties. The induced
gap states, being molecule mediated, depend on the length of the molecule, and on the 
interelectrode separation. The transport is determined by scattering states which are 
Bloch-like states in the electrodes and molecular orbitals like states
in between the electrodes.

Previous studies of the electronic transport of the junctions with the molecules of 
different length included a number of 
oligomeres in a search for a good molecular wire.\cite{JMT,Roncali}
For this purpose, one has to consider the molecules with
the same basic unit that repeats successively with
the length of the molecule.  The electronic structure at the
surface and the adsorption site geometry for that reason should be similar for all
the adsorbed molecules considered. Magoga and Joachim\cite{Mag} used a semi-empirical approach 
in their studies of the tunnel transport regime of certain families of oligomeres. 
They considered the transmission at zero bias voltage and found that the conductance 
follows a simple exponential decrease with the length of the molecule. 
However, in order to get a full account of the details of the 
electronic structure of the electrodes, the electrode surfaces, the molecule, and the 
molecule-induced gap states on the transport properties of the junction a 
full ab initio approach is needed.
Moreover, as the junction is subject to the non-equilibrium conditions
the transmission should be obtained self-consistently for each bias voltage applied across 
the junction.

In this paper, we therefore use the non-equilibrium Green functions technique and the density 
functional theory to obtain a full ab initio self-consistent description of 
the transport properties.
We investigate the conductance of the sulfur-ended
phenylene vynelene oligomeres coupled to gold electrodes; OPVn systems, with n equal to
the number of benzene rings in the molecule.
The number of benzene rings of these oligomeres determines the length of the molecule.
Owing to a strong
bonding of sulfur on metal surfaces, particularly gold, all other parameters
of the system can be kept fixed for the systems with different molecules of the series.
The difference in the transport properties of the coupled systems then reflects the
electrode distance in addition to the electronic structure of the molecule.
The results will show two different transport regimes\cite{my_talk}; the one where the 
electrodes are 
close enough that the induced gap states give an appreciable contribution to
the current, and the other where the
separation between the electrodes is large enough to result in their negligible contribution.

In Sec.~\ref{Calc} of the paper we first report on the model used and the generalities of the 
computational
procedure. In Sec.~\ref{Results} we calculate the current through the system as a function of
applied bias voltages and analyze them in terms of the evolution of transmission spectra.
 We discuss the results in Sec.~\ref{Disc}.

\section{\label{Calc} Calculations}

A typical molecular electronic system consists of a molecule coupled to two electrodes
with different electro-chemical potentials. Depending on the experimental method of 
building, the device may vary the contact geometry  with respect to the adsorption site as 
well as to the adsorption strength.\cite{new}
We consider an idealized geometry with a molecule
symmetrically coupled to the two identical gold electrodes.

\subsection{Computational method}
To perform the first-principle
quantum modeling of the electronic structure under non-equilibrium conditions and to
calculate the current-voltage (I-V) characteristics of the system,
we used non-equilibrium Green functions technique based on density functional theory
 as implemented in the recently developed TranSIESTA simulation package.\cite{Stokbro} 
Core electrons were modeled with
Troullier-Martins \cite{TM} soft norm-conserving pseudopotentials and the valence 
electrons were expanded in
a SIESTA localized basis set. \cite{SIESTA}
The system was divided into three regions: the left and right electrodes and the central region.
The central part contained the portion of physical electrodes, where all screening effects
took place. The
electronic occupations of the system was determined by the electrochemical
potentials of the electrodes. The charge distribution in the electrodes corresponded to the bulk
phases of the same material.

The density matrix of the system under the external bias was calculated self-consistently
within density functional theory (DFT) using the only major approximation of the method in
the choice of the exchange correlation functional. We used the local density
approximation (LDA). The virtue of the method is that it
gives the Hamiltonian in the same form as in the empirical tight-binding approach making 
its techniques straightforwardly applicable.
The nonlinear current through the contact is calculated using the 
Landauer formula\cite{Datta}

\begin{equation}
I(V_{b}) = G_{0}\int^{\mu_{R}}_{\mu_{L}}\, T(E,V_{b})\, dE,
\end{equation}

\noindent
where $G_{0}=2e^{2}/h$ is the quantum unit of conductance and  $\mu_{L/R}$ are
electrochemical potentials of the left and right electrodes. With the applied bias potential
$eV_{b}$, the left and right potentials become $\mu_{L}(V_{b})=\mu_{L}(0)+eV_{b}/2$ and
 $\mu_{R}(V_{b})=\mu_{R}(0)-eV_{b}/2$, respectively.  The energy region of the
transmission spectrum that contributes to the current integral in Eq.1. we refer to as
 the bias window. The total transmission
probability $T(E,V_{b})=\sum_{n=1}^{N}T_{n}(E,V_{b})$ for electrons incident
at an energy E through the device under the potential bias $V_{b}$ is composed of all
available conduction channels with the individual transmission $T_{n}$.

\subsection{OPVn molecules bonded to gold electrodes}
\begin{figure}[t]
\resizebox{1.0\columnwidth}{!}{
\includegraphics[clip=true]{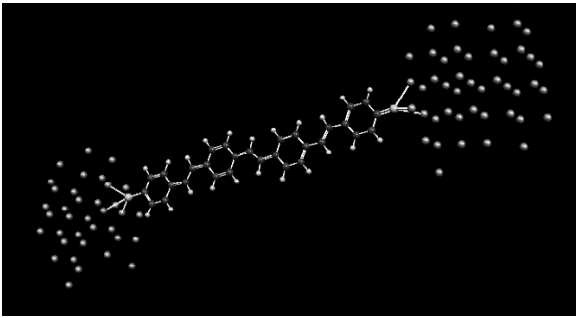}
}
\caption{\label{Au-opv4}
A phenylene vinylene oligomere, OPV4, connected to two Au(111) surfaces via thiolate bonds.
}
\end{figure}
The aim of this paper is to investigate the electrical and structural properties of 
the monolayer of
 phenylene vinylene oligomere molecules (OPVn) coupled to gold
electrodes via the sulfur atoms on both ends. There are good reasons to
believe that the result for the monolayer of OPVn molecules applies to the 
single molecule, too.\cite{single,Szuchmacher}

It is usually assumed that sulfur forms thiolate bonds
with the gold surface, which is not necessarily always the case.
Depending on the absorption technique, if the strong SH bond is not cleaved
and if the molecule forms a thiol bond with gold, a preferential adsorption site might
be at the top with the nonzero tilt angle. \cite{Walzer-Stokbro} In our calculation, however,
 we consider thiolate bonds at a 3-fold hollow site and the
tilt angle of the molecular axis to the surface normal to be zero.
We have also made the calculation with molecule adsorbed in the on-top position on the gold
atom at both electrodes, and found that the adsorption at the 3-fold hollow site is
energetically favorable to the on-top absorption by 2.6 eV. In the experimental situation,
 some molecules in the self-assembled monolayer will be atypical bonded to various kinds of 
defects on the surface.
We, however, consider an idealized situation and assume further that all
molecules 
align with parallel phenyl rings. There is a good reason to believe that this is
energetically more favorable, since it gives a larger $\pi$ overlap.

The geometry of the system was determined as follows.
We used a (3x3) unit cell of the Au(111) surface for the electrodes and assumed that
the molecule of cis conformation was chemisorbed to the surface of the electrodes, as shown
in Fig.\ref{Au-opv4}.
The OPV3 molecule was positioned symmetrically in the hollow site
of both Au(111) surfaces. After positioning the molecule perpendicularly in the 
z-direction
so that it has a favorable Au-S bonding distance,\cite{Andreoni} we let its molecular
coordinate to relax
until the average force on the molecular atom was less than 0.04 eV/\AA. The gold atoms
were kept fixed at their bulk positions. 

The same procedure was followed for OPV4 and OPV5
molecules. The obtained distances for Au-S atoms were $2.39 \AA$ for all systems
considered within the computational accuracy, the convergence in total energy tolerance
was 0.0001eV and the force tolerance was 0.04 eV/\AA .
In order to get an additional check on the stability of the results with respect to the
Au-S distance, we made a calculation for the OPV3 system with an Au-S distance enlarged
by 0.1 \AA.
The minor effects on the transmission with no consequences on the
current through the system were obtained. The lack of the force dependence on the
transport properties of the system with chemically bonded contacts 
was observed by Cui et al.\cite{Cui} on octaneditiol molecules.

In the TranSIESTA procedure the charge on the
molecule is not fixed. It adjusts itself to minimize the free
energy as the electrochemical potentials of the electrodes are changed. For the systems
considered, we have found that molecules preserve charge neutrality to within 0.15 e
with the change of the bias voltage.

The important aspect of the calculation is the robustness of the results to
computational details.
From the calculation by Stokbro et al.\cite{Stokbro} for the dithiolate benzene molecule
on Au(111) surfaces (dtb) in the hollow position we know that the use of the generalized
gradient approximation (GGA) instead of the LDA gives almost negligible
effects on the transmission spectra. The unimportance of the 
relaxation of the first two gold layers for final results was also demonstrated. 
As the binding geometry is
mostly determined in the first two layers of gold atoms, the conclusion is 
valid for the systems with OPV3, OPV4 and OPV5 molecules as well.

\section{\label{Results} Transmission properties and I-V characteristics}

In order to obtain I-V spectra and differential conductance, we performed
self-consistent calculations for a bias voltage in the range from -3.0 V to 3.0 V.
Since our electronic
structure in the contact region is fully symmetric, the corresponding I-V spectra are
symmetric with respect to the reversed bias.

\begin{figure}[]
\resizebox{0.9\columnwidth}{!}{
\includegraphics[clip=true]{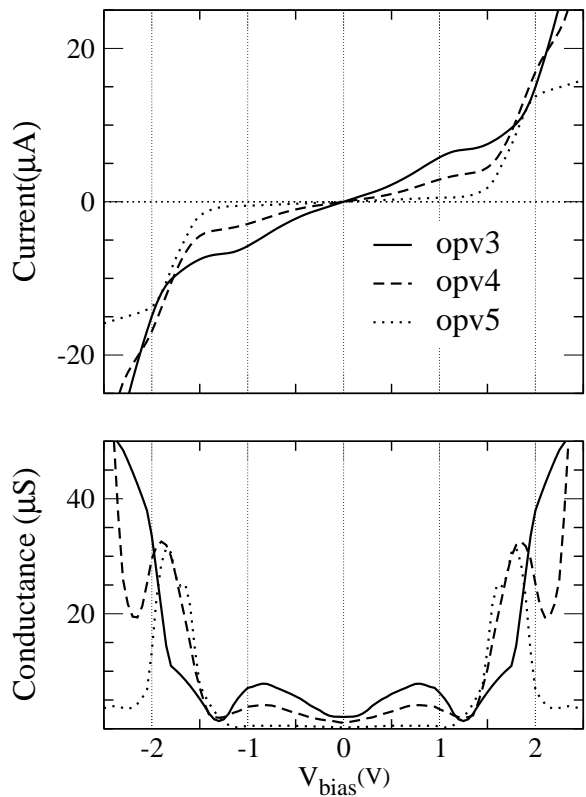}
}
\caption{\label{I-V}
Current and differential conductance of OPVn systems as a function of bias. A self-consistent
calculation has been employed for each bias voltage.
}
\end{figure}

The general trend of I-V curves is  shown in Fig.\ref{I-V}. One observes that all
 curves
have a similar general shape. They show a slow increase of the current
with the bias voltage up to a certain bias and then a steep increase at a point where
transmission resonances come into alignment with the bias window.

Two characteric features of the systems can be observed.
The first one is that the current in the ohmic region of the transport is lower as
the length of the molecule is larger.
The second one is that a steep increase of the current of the longer molecule happens
at smaller bias than for the short molecule.

When compared to the results for shorter DTB system\cite{dtb_Au} the absolute value 
of the current of the OPV3 is almost an order of magnitude smaller.
It is also in good agreement with the current of the similar type of 
molecules.\cite{Reichert,Heurich}

The zero bias conductance $G\!=\!\frac{e^2}{h}\!T(\mu_{L/R},0)$ is also
much smaller than for DTB. They are 2.1 $\mu S$ for OPV3, 
1.0 $\mu S$ for OPV4, 0.2 $\mu S$ for OPV5, in comparison with 35.0 $\mu S$ for DTB.

In the following we analyze the obtained IVCs for OPV3-5 in terms of the dependence
 of the transition
spectra on molecular length (number of benzene rings), bias voltage, and 
interelectrode separation.

\subsection{OPV3}

Chemisorption of the molecule on the gold surfaces leads to the change of the energy
levels in both systems. In particular, some of the molecular levels broaden into a continuum.
The eigenstates of the combined system become the scattering states which
are Bloch-wavelike
in the electrode and  molecular orbital-like in the molecule.\cite{Taylor}
The more delocalized the state is, more spreading the corresponding transmission peak shows.

\begin{figure}[h]
\resizebox{0.95\columnwidth}{!}{
\includegraphics[clip=true]{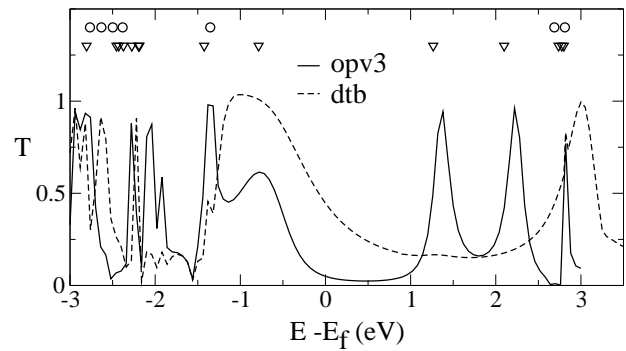}
}
\caption{\label{dtb,opv3}
Transmission amplitude of DTB and OPV3 molecules
on Au(111) surfaces at zero bias voltage. MPSH eigenvalues are marked with circles for
dtb and with triangles for OPV3.
}
\end{figure}

\begin{figure}[h]
\resizebox{0.95\columnwidth}{8cm}{
\includegraphics[clip=true]{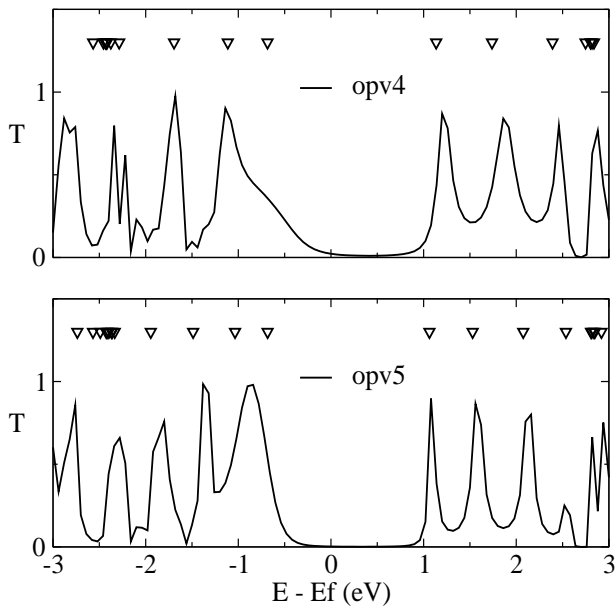}
}
\caption{\label{opv3,4,5}
Transmission amplitude and MPSH eigenvalues of OPV4 and OPV5 systems at zero bias.
}
\end{figure}

In Fig.\ref{dtb,opv3} we compare the transmission spectra of the coupled dtb
system with the OPV3 system at zero bias.
In the figure we also mark the eigenvalues of the projected self-consistent
Hamiltonian (MPSH) onto the molecular orbitals.
The transmission spectra are proportional to the density of states of the combined
system. The MPSH levels are normally broadened and shifted in energy by the interaction with
the gold surface. Some of the states have a large weight on the S atom and form a strong
 bond
with the gold electrode. The others may have a little overlap with Au states and form
a narrow peak in the density of states and in the transmission spectra.

In the DTB system the huge transmission resonance below the Fermi level is due to
a superposition of strongly hybridized levels,
as demonstrated in Ref.~\onlinecite{dtb_Au}.

The OPV3 system does not exhibit such a huge resonance as DTB does and it has
a considerably less spectral weight at the Fermi level.
The OPV3 resonance below the Fermi level
is built on the HOMO  molecular state at the energy $E\!=\!-0.799eV$.
The reason for being less pronounced than in DTB is the 
larger separation of Au electrodes in OPV3, resulting in the smaller overall
overlap of gold states with the molecular states.
This leads to smaller hybridization and less broadened levels resulting in lower
density of states around the Fermi level. 

\begin{figure}[h]
\resizebox{0.95\columnwidth}{!}{
\includegraphics[clip=true]{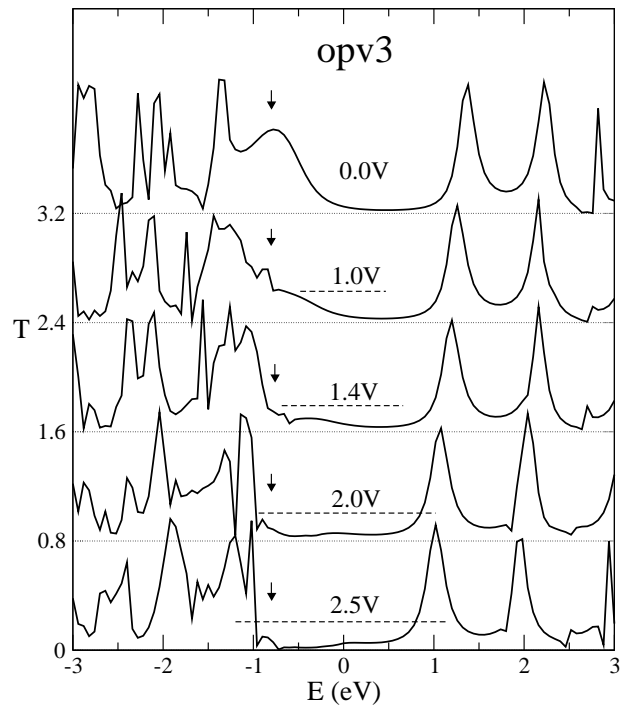}
}
\caption{\label{opv3-cent}
Transmission $T(E,V_{b})$ of the OPV3 system as a function of the bias voltage (shifted
vertically by 0.8 for visibility). Energies are relative to the average electrochemical potential
$(\mu_{L}+\mu_{R})/2$. The bias windows are marked with dashed lines. The amplitude
corresponding to the HOMO levels (marked with arrows) decrease with the increased bias.
}
\end{figure}

When the bias $V_{b}$ is applied to the system the chemical potential of the electrodes
changes. The calculated transmission spectra are
 plotted for a set of bias values in Fig.\ref{opv3-cent}.
 The main effect of the increased bias is that more of the resonance gets into the bias
window.
The position of the HOMO related transmission resonance does not change with respect to the 
average electrochemical potential as the bias is
 increased, but its weigth diminishes.
The resonance gives a
lower contribution to the current tranport at higher bias, even when the maximum of
 the resonance enters the bias window. That peculiar bias dependence of the HOMO resonance
leads to a step-like behavior of the current and a dip in the conductance 
at the bias of $1.25V$ as seen in Fig.\ref{I-V}.

The transmission peaks corresponding to LUMO,
and higher levels change linearly with the applied bias, contrary to the HOMO peak.
 They are slowly approaching the
 electrochemical potential of the left electrode leading to a smaller 
HOMO-LUMO separation. The net result is the lower transmission in the 
HOMO-LUMO region. The current is considerably smaller 
when compared to the current of the smaller dtb system, 
particularly at low bias. Eventually, at high enough bias, at above approximately 2V the LUMO
resonance enters significantly the bias window and contributes dominantly to the current.
This is the region of the steep increase of the current as seen in Fig.\ref{I-V}.
Nevertheless the current in the OPV3 system is almost an order of magnitude
smaller than in the DTB system in the region we considered.

\subsection{OPV4}
\begin{figure}[h]
\resizebox{0.95\columnwidth}{!}{
\includegraphics[clip=true]{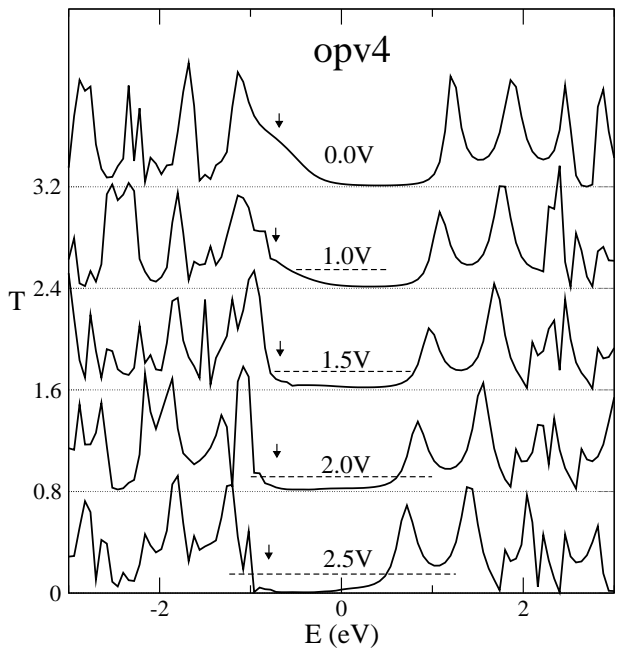}
}
\caption{\label{opv4,Vb}
Transmission $T(E,V_{b})$ of the OPV4 system for different bias voltages
The notation is the same as in Fig.\ref{opv3-cent}. The HOMO amplitude
decreases with increased bias and the LUMO amplitude moves towards the average 
electrochemical potential of the electrodes.
}
\end{figure}

 There is a clear separation of
 HOMO and LUMO levels in the OPV4 system for zero bias, as shown in Fig.\ref{opv3,4,5}.
The appearance of the lower transmission of the HOMO related resonance and the lower 
density of states at $E_{f}$ is  more
pronounced than in OPV3 system. The position of the HOMO level is $0.685 eV$
below the Fermi level. The LUMO level is $1.136 eV$ above the Fermi level
and is less broaden giving the lower tail at the Fermi level.
The behavior is maintained at the finite bias, as seen in Fig.\ref{opv4,Vb}.
The MPSH levels are narrower in
general, and the resulting transmission at low bias is smaller when compared 
with the OPV3 system as seen in Fig.\ref{I-V}.
Like in the OPV3 a dip in the conductance appears at $1.25V$.

\subsection{OPV5}

The tendency of the appearance of a sharper well-defined peaks in the system with the longer
molecule is clearly demonstrated in the OPV5 system in Fig.\ref{opv3,4,5}. The contribution of 
the
underlying states to the resonance is obviously changed with respect to the OPV4. The overall
overlap of the molecular states with the gold electrode states is lower leading to the narrowing
of the transmission resonances.
The HOMO and LUMO resonances at zero bias are at $0.684 eV$ below and $1.063 eV$ above the
Fermi level, respectively.
With the increased bias, the HOMO transmission almost disappears, while the LUMO slightly
moves towards the average chemical potential of the electrodes.
The dip in the conductance at $1.25V$, which was present in OPV3 and OPV4, is missing 
in OPV5 owing to a very little spectral weigth of the HOMO resonance.

\begin{figure}[h]
\resizebox{0.95\columnwidth}{!}{
\includegraphics[clip=true]{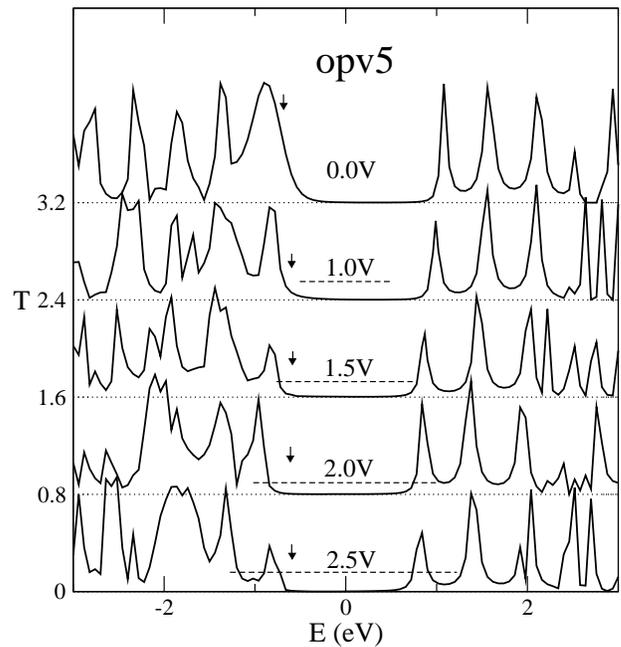}
}
\caption{\label{opv5,bias}
Transmission $T(E,V_{b})$ of the OPV5 system. The notation is the same as in Fig.\ref{opv3-cent}
The behavior of the HOMO and LUMO amplitudes
is similar to that of the OPV3 and OPV4 systems. Note a low density in the HOMO-LUMO region.
}
\end{figure}

There is a new feature in the spectra related to the
narrowing of the levels.
 An almost pure HOMO-LUMO gap sets up at any bias.
The bias dependence of the gap is reflected only through its slight shift
with respect to the average chemical potential of the electrodes, as seen 
in Fig.\ref{opv5,bias}. The  low zero-bias conductance is 0.2 $\mu S$, and the current is 
distinctly lower compared with the preceding member of the OPVn series.

\section{\label{Disc}Discussion}

The current-voltage dependence of the systems can be obviously understood in terms of
the transmission resonances.
The OPV3 system at zero bias does not show a huge transmission resonance
below the Fermi level, as was characteristic of a smaller DTB system.
The tendency of diminishing transmission and density of state around the Fermi level
is more pronounced for the OPV4 system and even more for the OPV5. Loosely speaking, for a short
molecule there will be a strong molecule-mediated overlap of electrode states leading to wide 
transmission resonances as opposed to the weaker overlap, and more narrow transmission
peaks for longer molecules.

To understand this consideration better, let us consider the transmission related to the 
interelectrode
separation. We define it as the distance between the surface planes of the left and 
right electrodes.
In our calculations the shortest distance is in the DTB system with the value
1.05 nm. Our input radius of the gold atoms is 3 \AA, and the direct overlap of the gold atom 
states from the left and the right electrodes cannot take place.
The pure gold levels, which perfectly follow the electrode chemical potential,
can contribute directly to the tunneling process through the electrode-induced resonant states
or even quantum well states related to gold surface states.\cite{Ghosh}
The properties of these states are dependent on the separation between the electrodes.
As reported by Kergueris et al.,\cite{Kerg} the tunneling conductance in the gold-air-gold
junction is linear and of the order
0.02 $\mu S$ at the barrier height of 1V for the interelectrode spacing of 1 nm.
That appears to be several orders of magnitude
smaller than the typical conductance of metal-molecule-metal junctions.
However, with the molecule chemisorbed in between the electrodes, the electrode induced gap
states and resonances play an important role in transport properties.
 They are, of course,
a part of the hybridized molecule-electrode states, caused by the molecule-mediated
overlap of electrode states. We can tentatively call them
molecule-induced gap states.

If the electrodes are close, they contribute to the finite density
of states and to the enhanced transmission in the HOMO-LUMO region of the combined
systems. As the electrode separation is larger, the molecule-induced
gap states loose the weight,
and their contribution to the conducting channels of the system becomes smaller.
The transmission
peaks become narrower and the transmission in the HOMO-LUMO region
consequently smaller.

The DTB molecule is the shortest one we have considered.
The increase of the interelectrode separation in the OPVn series from 2.3 nm for OPV3
to 3.64 nm for OPV5 is followed by the decrease of
the zero-bias conductance, as seen in Table~\ref{distance}\cite{comment}.

\begin{table}[]
\caption{\label{distance}
Interelectrode separation and zero-bias conductance of OPVn systems.
$d_{elec.}$ is the distance between the surface
(111) gold planes of the left and right electrodes. G($\mu_{L/R}$, $0$) is
the conductance at the Fermi level for zero bias.
}
\begin{ruledtabular}
\begin{tabular}{ccccc}
  & DTB & OPV3 & OPV4 & OPV5  \\ \hline
      \cline{2-5}

$d_{elec.}$ & 1.05 nm & 2.3 nm & 2.97 nm & 3.64 nm\\
G($\mu_{L/R}$, $0$) & 35 $\mu S$ & 2.1 $\mu S$ & 1.0 $\mu S$ & 0.2 $\mu S$

\end{tabular}
\end{ruledtabular}
\end{table}

\begin{figure}[h]
\resizebox{0.9\columnwidth}{4.5cm}{
\includegraphics[clip=true]{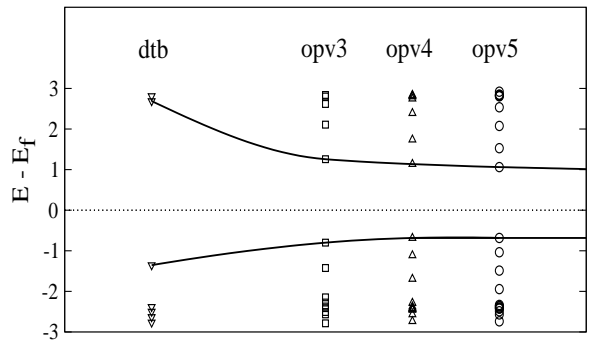}
}
\caption{\label{mpsheig}
Eigenvalues of the projected self-consistent Hamiltonian of the Au(111)-molecule-Au(111)
 system onto the molecular orbitals at zero-bias.
Full lines connect the HOMO and LUMO levels, respectively.
}
\end{figure}

Indeed, as discussed in Section~\ref{Results}, the corresponding transmission in the HOMO-LUMO
region of those junctions systematically drops down with the distance
between the electrodes, as seen in Figs.\ref{dtb,opv3} and \ref{opv3,4,5}. The true discrete
nature of molecular resonances starts to govern the transport.
The level spacing of the resonances is illustrated in terms of the eigenvalues of the
projected self-consistent
Hamiltonian onto the molecular orbitals (MPSH) in Fig.\ref{mpsheig}.
The levels are more closely spaced and the broadening of the levels goes down with
the length of the molecule.

The semiconducting property with a narrow band gap of the $\pi$ conjugated system is well known.
The gap is due to a distinct alternation of the bonds, with every second bond having
double-bond character.
In the poly p-phenylene vinylene molecule, the HOMO-LUMO gap of $2.5 eV$ has been
reported.\cite{Friend} The reduction of the gap obtained in this paper, 
 as seen in Fig.\ref{mpsheig}, is expected for the 
thiolate-ended molecule on gold surfaces.

Another characteristic feature of the OPVn series is its zero bias conductance.
One finds a remarkable difference between the DTB and OPV5
systems. For the DTB, it is $35 \mu S$, while for the OPV5 of the same contact geometry it is
only $0.2 \mu S$. 
Obviously, the transmission in the HOMO-LUMO region appreciably drops down in
the series as seen from Fig.\ref{opv3-cent}-\ref{opv5,bias}.
For an even longer molecule, one expects a negligible density of states 
around $E_{f}$ and a vanishing transmission in the HOMO-LUMO region for low bias voltages.

These two features, the fixed HOMO-LUMO separation and the gap in the density of states around
$E_{f}$, show that a kind of bias threshold sets up. A current through the
junction will start at a bias above some threshold value, at which the wings of the HOMO and LUMO
peaks enter the bias window. This is a kind of  bias-activated coherent transport.

One expects the gap in the I-V curve to be more pronounced in the systems with weaker 
chemisorptions than in the sulfur-ended molecule on gold electrodes.

Owing to the bias gap, a long  molecule, even though
chemisorbed on the electrodes, may show a behaviour of a
small capacitor.\cite{Mujica,Kerg,Zhou}  For nonequal bonding to the electrodes, this
quite often happens in real junctions. This opens the
possibility for a kind of sequential transport in $\pi$ conjugated systems and charging
of the strongly chemisorbed molecule in gate-controlled experiments.
Kubatkin et al.\cite{Kubatkin} have demonstrated gate-controlled Coulomb blockade involving
different charge states in the OPV5 molecule of tran conformation, but with 
the protected thiol end-groups of their molecule, thus preventing strong
chemisorption.

In the view of this, a long molecule coupled to the electrodes may also serve as a testing
ground for the
boundary region between mesoscopic and nanostructure systems.

We also point out that
the calculations we have performed showed that the hollow position is the minimum-energy 
absorption site for all molecules in the series. We have taken the
 distance of sulfur to gold atoms as $5\% $
 larger and found that it
does not change the conduction noticeably.\cite{pressure} Besides the known mechanical
and temperature stability of thiolate-ended organic
molecules bonded to gold surfaces,\cite{Reed, Kerg} this robustness of conduction is an
additional desirable property in
the view of applicability in molecular electronics.

\section{\label{conclusion}Concluding remarks}

Using the NEGF and DFT we have obtained the nonlinear current-voltage
characteristics of
thiolate-ended phenylene vinylene oligomers coupled to Au(111) electrodes. In the low-bias
region we have found that the current scales down inversely to the interelectrode separation,
with the strong tendency of the appearance of the I-V gap for long molecules.
The reason for the gap formation is the narrowing of the HOMO and LUMO related transmission peaks with
the interelectrode separation. The nonexistence of the I-V gap in the junctions
of short molecules appears to be a consequence of the ideal Au(111) semi-infinite
surfaces and the molecule-mediated gap states encountered in the calculations 
resulting in wide
resonances and finite density of states around the Fermi level.
The measured conductance of short molecules is usually smaller than
the calculated one, which is possibly due to such ideal conditions seldom, if ever, encountered in
real junctions. For longer molecule, the effects appear to be less efficient, and as expected
the discrepancies between the measured and calculated values are less pronounced.
We point out that the full account of such effects could only be followed with the full ab initio
self-consistent calculations we performed at each bias voltage.

We have also found that the conductance is robust to the sulfur gold distance for the
molecules chemisorbed in the local energy minimum, hollow site on gold surfaces.


\begin{acknowledgments}
This work was supported in part by the Ministry of Science and Technology
of the Republic of Croatia under contract No. 0098001, by the IST-FET-NANOMOL
project of the EC, by the Swedish Research Council and by the Swedish
Foundation for Strategic Research.

\end{acknowledgments}


\begin{thebibliography}{00}
\bibitem{SIESTA} A double $\zeta +$ polarization basis for the OPVn molecule,
while a double $\zeta$ for the gold {\it s} channel and single $\zeta$ for the gold
{\it p} and {\it d} channel was used.
\bibitem{comment} Our calculations do not show an exponential decrease of 
the conductance with the interelectrode separation for the 
lengths we have considered. That is a new feature with respect to the  
the semi-empirical approach of Ref.~\onlinecite{Mag}, showing that the full ab initio
approach should be used for the junctions with chemisorbed molecules.
\end{thebibliography}

\end{document}